\documentclass[a4paper, 10pt, twocolumn]{article}

\usepackage[dvips]{epsfig}
\usepackage{rotating}
\usepackage{graphics}
\usepackage[dvips]{color}

\begin{document}
\begin{titlepage}

\begin{flushright}Edinburgh 2000-27 \\ 
Dec 2000
\end{flushright}

\vskip 1.0cm
\large
\centerline {\bf Status of the LHCb Experiment\footnote{This 
presentation is dedicated to Tom Ypsilantis,  1928 - 2000.}
}
\normalsize
 
\vskip 2.0cm
\centerline {Franz Muheim\footnote{E-mail: F.Muheim@ed.ac.uk} }
\centerline {on behalf of the LHCb Collaboration }
\centerline {\it {Department of Physics and Astronomy, The University of Edinburgh,} }
\centerline {\it {Mayfield Road, Edinburgh EH9 3JZ, Scotland/UK} }

\vskip 4.0cm
 
\centerline {\bf Abstract}
\vskip 1.0cm
We present the status of the LHCb experiment which will make precision
measurements of CP violation in $B$ meson decays.
The motivation for the experiment and an overview of the detector design are given.
  The vertex detector, ring imaging Cherenkov counter, 
calorimeters, and trigger systems are discussed in
detail. We also present the expected physics performance for 
selected modes.
\vskip 2.0cm
\centerline {\it Presented at 7th International Conference on B-Physics at Hadron} 
\centerline {\it Machines, (BEAUTY2000), Sep 2000, Kibbutz Maagan, Israel.}
\centerline {\it Submitted to Nuclear Instruments and Methods A.}
\vfill
\end{titlepage}

\section{Introduction and Overview }

\begin{figure}[!t]
\begin{center}
\mbox{\epsfig{figure=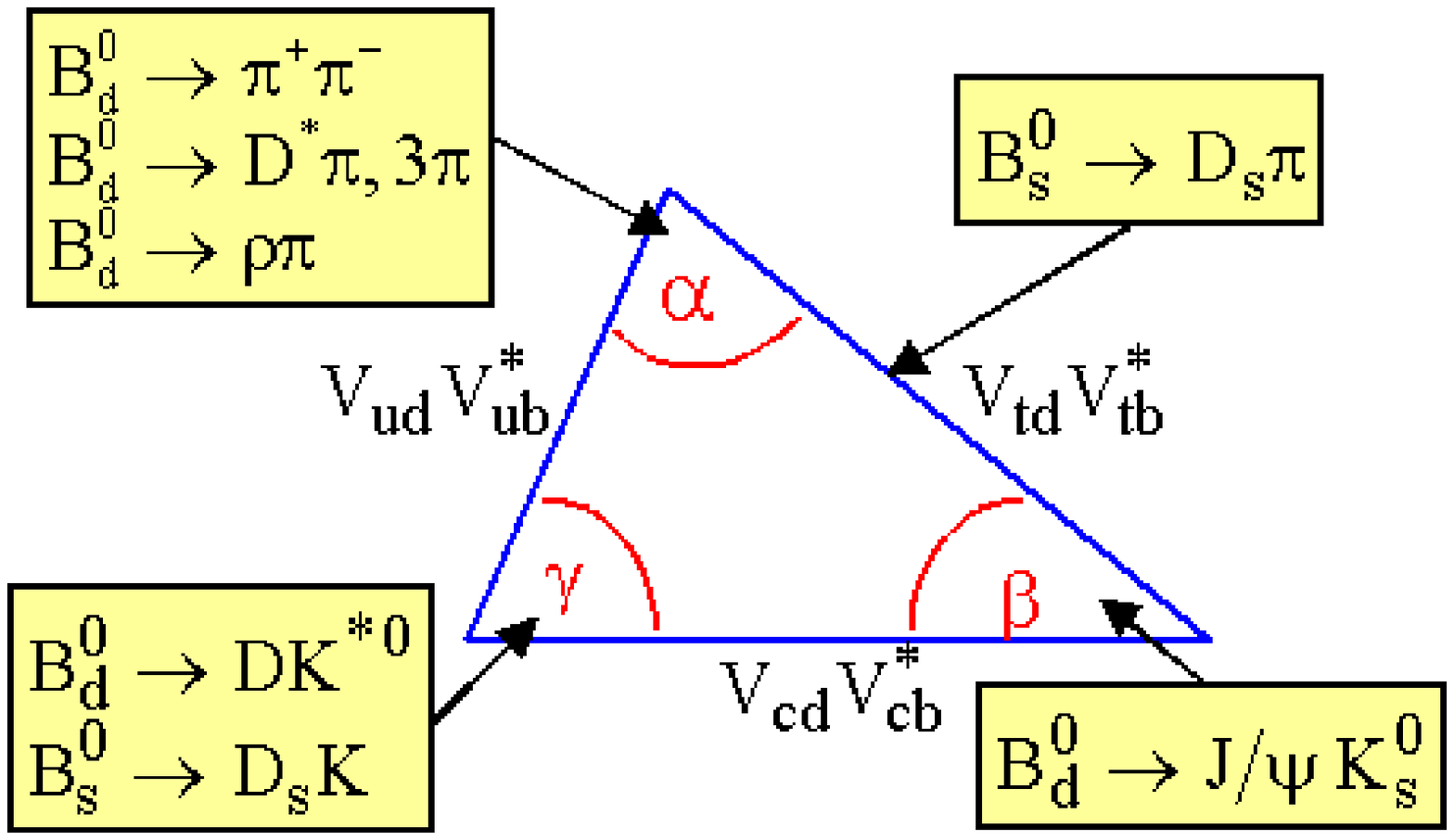,width=0.5\textwidth}}
\mbox{\epsfig{figure=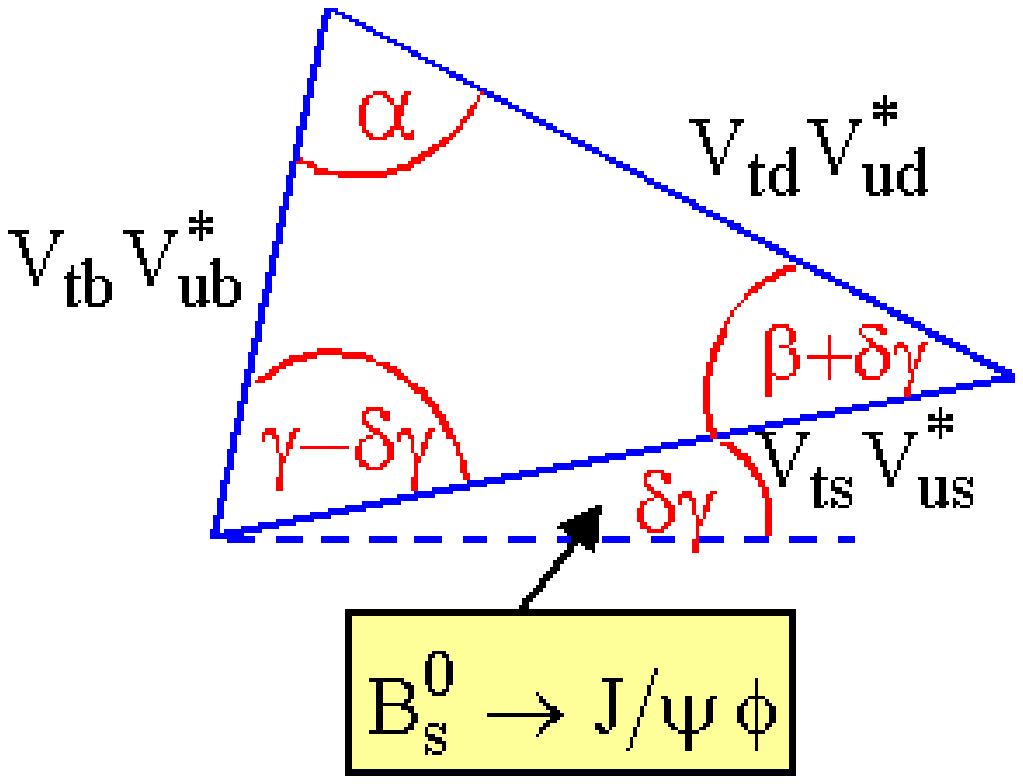,width=0.4\textwidth}}
\vspace{-0.2cm}
\caption{Two unitarity triangles. B decays and the angles and sides
which they probe are also given.}
\vspace*{-0.5cm}
\label{fig:uni}
\end{center}
\end{figure}

CP Violation is one of the most outstanding puzzles in
particle physics. More than
third of a century after its discovery,  CP violation still has only been observed
in the neutral kaon system. In the Standard Model (SM) of particle physics
the mixing of quarks is
described by the Cabibbo-Kobayashi-Maskawa (CKM) matrix. 
For three generations of quarks and leptons, 
the CKM matrix has four free parameters,
three Euler angles, and one complex phase. 
It is this phase which allows CP violation to occur within the SM.
While CP violation in the kaon system is tiny and
plagued by theoretical uncertainties
the SM predicts large  CP violating effects in the B meson system.
This is illustrated by the two relevant 
unitarity conditions
\begin{eqnarray}
 V_{ud}V^*_{ub}+V_{cd}V^*_{cb}+V_{td}V^*_{tb}&=&0 
\nonumber \\
 V_{tb}V^*_{ub}+V_{ts}V^*_{us}+V_{td}V^*_{ud} &=&0
\nonumber
\end{eqnarray}
which can be represented by unitarity triangles in the complex plane
as shown in Figure~\ref{fig:uni}. 
The lengths of the sides are all of the same order of magnitude.
CP violating asymmetries correspond to  non-zero values for
the angles $\alpha$, $\beta$, and $\gamma$.
The relevant B decay channels are also indicated.
%
%
%

However, we have no real understanding of CP violation. For example, baryogenesis 
tells us that beside 
the SM additional sources of CP violation are needed.
For B mesons  the SM makes accurate predictions in many decay channels
which can be tested. Precise measurements of many CP violating angles
overconstrain the unitarity triangles and
allow to extract the  parameters of the SM and of the possible new physics.

By 2005 the experiments {\sc BaBar}, BELLE, CLEO-III, CDF, D0, and HERA-B
will most 
likely have observed CP violation in the channel $\mathrm{B \to J/\psi K^0_S}$
which determines the angle $\beta$.
LHCb is a next generation experiment,
starting in 2005 at the Large Hadron Collider (LHC).
Given the large cross section of 500~$\mu$b
the LHC will deliver $10^{12} \; b \bar b$-quark pairs
per year  at a modest luminosity of 
$2 \times 10^{32} \rm cm^{-2} \rm s^{-1}$.
All types of B mesons ($\mathrm{B_d}$, $\mathrm{B_u}$, $\mathrm{B_s}$, $\mathrm{B_c}$)
will be produced. 
The LHCb experiment 
is designed to make
precision measurements of CP violation in $\mathrm{B}$ mesons 
and to overconstrain the CKM matrix.

\begin{figure*}[htb]
\begin{center}
\mbox{\epsfig{figure=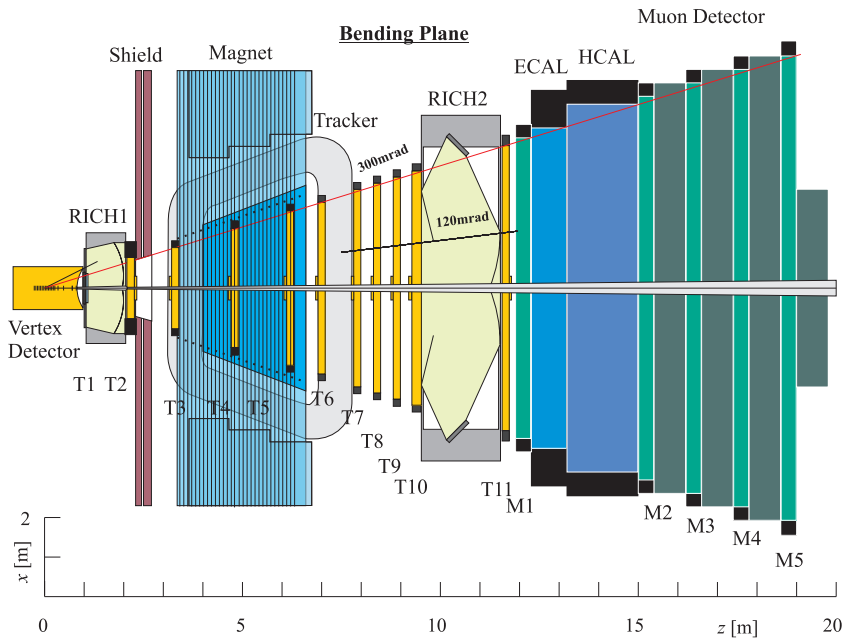,width=0.7\textwidth}}
\caption{The LHCb spectrometer seen from above (bending plane).}
\label{fig:lhcb}
\vspace{-0.5cm}
\end{center}
\end{figure*}
The LHCb detector shown in Figure~\ref{fig:lhcb}
 is a forward single arm spectrometer which exploits 
the fact that  $b$ and the $\bar b$ quark production
angles are correlated and  
and that the production 
rates  are strongly peaked forwards and backwards.
The main challenges of the experiment are: i) the trigger which must select
leptonic and hadronic final states, for example
$\mathrm {B^0_d \, \rightarrow \,\pi^+ \pi^-}$,  out of the much more abundant
background; ii) charged particle identification, i.e.
the ability to distinguish between pions and kaons  over a large  momentum range 
and iii) an excellent proper time resolution 
for secondary vertices 
to separate $\mathrm {B}$ meson decays from light quark decays.

The detector design is as follows:
A spectrometer magnet consisting  of a normal conducting
coil provides a bending power of $\mathrm{\int{B dl} = 4 \; Tm}$
with the field oriented vertically.
The detector acceptance is $\pm 300$~mrad ($\pm 250$~mrad) in the
(non-) bending plane.  
Charged tracks are reconstructed in the inner and  outer tracker.
The inner tracker technology is either 
a triple-GEM gaseous or a silicon strip detector. 
The outer tracker consists of straw tube drift chambers, and is
discussed in a separate contribution~\cite{ref:outer}. 
Wheels of silicon strip detectors are located around the interaction point
for the reconstruction of primary and secondary vertices.
Charged particle identification is provided by 
Ring Imaging Cherenkov (RICH) counters. 
The electromagnetic and  
hadronic calorimeters 
are used for triggering and shower reconstruction.
The muon system is described in detail in another 
contribution~\cite{ref:muon}. The trigger is divided into four
levels and the LHCb computing strategy is discussed in 
a separate contribution~\cite{ref:comp}. 

After the submission of the Technical Proposal~\cite{ref:TP}
the LHCb experiment was approved in Sep 1998. Since then many major technology
decision have been made. The Technical Design Report (TDR) for the Magnet has
been approved in April 2000~\cite{ref:magnet_tdr} 
and very recently, in September 2000, the RICH and the 
Calorimeter subsystems have submitted their 
TDRs~\cite{ref:rich_tdr,ref:calorimeter_tdr}.   

\section{Vertex Detector}

The choice of technology for the LHCb vertex detector 
is single sided p-n or n-n silicon strip detectors with double metal read-out.
The design of the  vertex locator (VELO) is optimised for the  Level-L1 trigger.
Thus $r$ and $\varphi$ strip detectors alternate and the strip pitch  varies
such that the occupancy is roughly constant and below 0.5~\%.
The VELO wheels are split in halves
and will be mounted onto radio frequency (RF) shields
situated inside  the vertex tank. 
During injection of beams into LHC 
the detector halves will be retracted by 3~cm which
allows an active inner radius of 8~mm during physics runs.
Depending on the actual radiation dose the  detectors will be exposed to
they may have to be replaced after a few years running of LHCb.

An optimisation of the VELO design has been carried out 
at the MAP simulation farm based on 300 Linux PCs~\cite{ref:MAP}
by generating, reconstructing, and analysing over  10 million events.
The major  changes of this improved VELO design are 
a ``toblerone''-like shaped RF shield, an increase in the number
of stations from 17 to 25 and 250~$\mu$m thick detectors with
smaller pitch and radii. 
The  design has an excellent
proper time resolution of 43~ps for the decay 
$\mathrm {B^0_s \, \rightarrow \, D_s^{\mp} \pi^{\pm}}$ which corresponds
to a sensitivity of up to $\mathrm {x_s \sim 75}$ for  
$\mathrm {B^0_s \bar B^0_s}$ oscillations.
In addition, 9.5 hits per track  are expected on average
which allows for stand-alone tracking in the VELO.
\begin{figure}[thbp]
\begin{center}
\mbox{\epsfig{figure=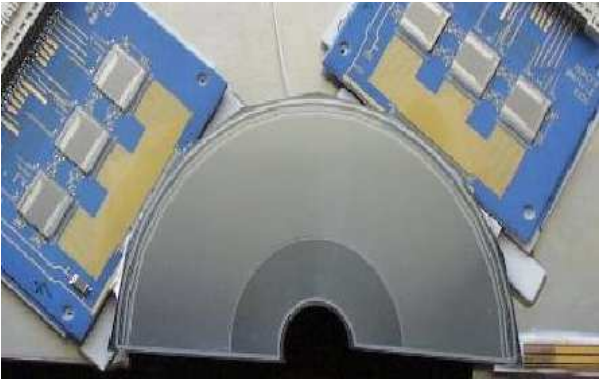,width=0.5\textwidth}}
\caption{A silicon prototype detector equipped with LHC speed electronics.}
\label{fig:velo-setup}
\vspace{-1.0cm}
\end{center}
\end{figure}

Two designs for the front-end electronics are currently 
under evaluation a) the SCTA128\_VELO chip a prototype version 
of which has been used for testing the silicon detector 
and b) the BEETLE chip will
be examined as soon as it becomes available. 
The design of a first prototype board of 
the off-detector electronics has been completed
and tests are ongoing.

An extensive R\&D programme to test the characteristics
of the vertex detector is being carried out. 
Prototype silicon detectors 
equipped with  LHC speed electronics, shown in Figure~\ref{fig:velo-setup},
have been successfully operated in beam tests at CERN.
We have measured the signal shape,
the spatial resolution and the cluster finding efficiency. 
Prototype detectors have also been heavily irradiated to simulate the 
expected radiation dose of more than three years of LHCb running.
The performance before and after irradiation has  been 
measured.
    
In summary the VELO design is well matched to   the 
criteria important for vertexing in LHCb. The geometry with $r$ and $\varphi$ strips
is optimum for the Level-L1 trigger. The pitch is smallest at the inner
radius which improves upon the  impact parameter resolution. 
The detectors are thin which minimises multiple scattering
and the relatively low number of channels reflects in the cost of the system.
 
\section{Ring Imaging Cherenkov Counters}

The ability to distinguish between pions and kaons in various final states
is a fundamental requirement of LHCb. In many important 
$\mathrm{B} $-decay channels CP violation measurements are only possible
with excellent hadron identification. For example, there are
three kaons in the final state of the decay chain 
$\mathrm {B^0_s \, \rightarrow \, D_s^{\mp} K^{\pm}}$,
$\mathrm {D^-_s \, \rightarrow \, \varphi \pi^{-}}$,
$\mathrm {\varphi \, \rightarrow \, K^+ K^-}$.
The momentum spectrum of these tracks is quite broad 
extending beyond 100 GeV/c.
Identifying kaons from the accompanying B hadron decay in the event
of which a large fraction have momenta below p $<$ 10 GeV/c
also provides a valuable tag of the $b$ quark production flavour.
This means that the  $\mathrm{\pi - K}$ separation must span the range from 
$\mathrm{ 1 \; GeV/c < p < 150\; GeV/c}$.

\begin{figure}[htbp]
\begin{center}
\mbox{\epsfig{figure=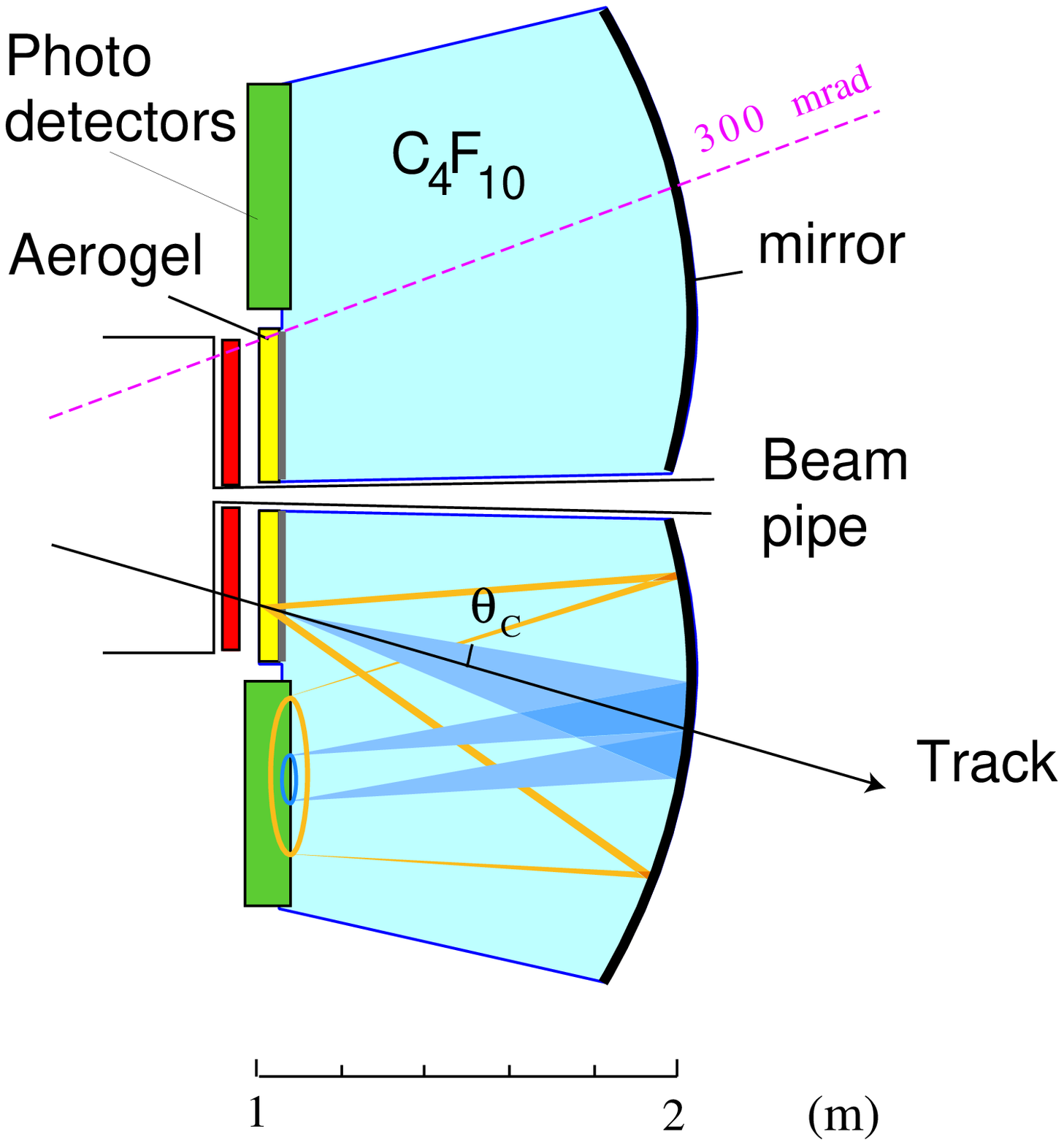,width=0.25\textwidth}
        \epsfig{figure=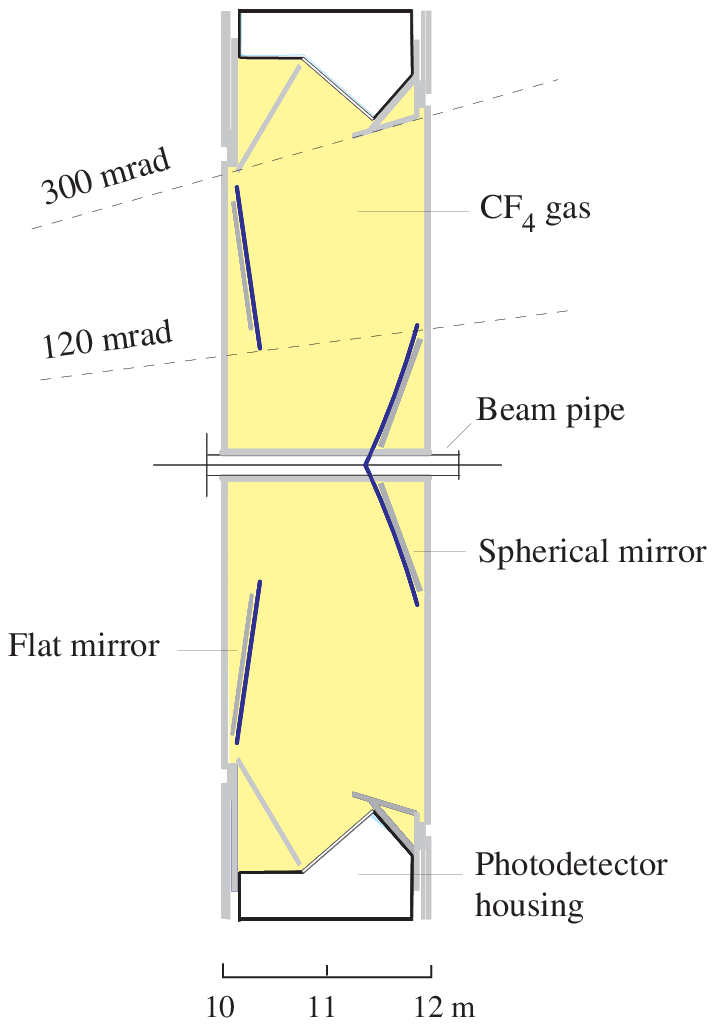,width=0.25\textwidth}}
\caption{Layout of the RICH\,1 and RICH\,2 detectors.
The production of two rings in the photo detectors is illustrated for 
RICH\,1. }
\label{fig:rich}
\end{center}
\end{figure}

This performance can only  be achieved with RICH detectors.
There is a strong correlation between the momentum and the polar angle of
tracks, at wide angles the momentum spectrum is softer. 
Hence the RICH system is divided into two detectors shown
schematically in Figure~\ref{fig:rich}. 
An upstream detector (RICH\,1) covers the full acceptance of
LHCb. RICH\,1  contains both 
a Silica aerogel radiator with $n = 1.03$ which is suitable for  
the lowest momentum tracks and a 
gaseous $\rm{C}_4\rm{F}_{10}$ radiator well matched for 
the intermediate momentum region.
A downstream detector (RICH\,2) has a $\rm{CF}_4$ radiator, to analyse
 high-momentum tracks. Its coverage is
limited to the region 120\,mrad (horizontal)$\times 100$\,mrad (vertical). 
The three radiators are required to cover the full momentum range.
In both detectors the Cherenkov photons are focused by mirrors onto 
photo detector planes which are positioned outside the LHCb acceptance.

An intensive research and development programme has been undertaken
for the LHCb RICH detectors.
A major effort 
has been the development of the photo detector system,
which must cover an area of $2.6 \rm{m}^2$ 
with a large active area fraction at an acceptable cost.  
The RICH photo detectors must be sensitive to single photons with a
peak quantum efficiency  larger than 20~\%
and provide a spatial granularity of $2.5 \times 2.5 \;  \rm{mm}^2$.
The read-out electronics for the photo detectors 
must be compatible with the LHC bunch crossing
frequency (40~MHz) and the devices must work in an environment with 
large charged particle fluxes and fringe magnetic fields of up to 5 mT.  
Three options were investigated for the photo detectors: 
two based on hybrid photo diodes (HPD), 
the Pad HPD and the Pixel HPD; and 
the 64 channel Multianode Photo Multiplier Tube (MaPMT).
A panel reviewed the three options and subsequently
the Pixel HPD has been selected by the LHC collaboration as baseline
photon detector. 
The MaPMT is kept as a backup option
and the Pixel HPD must meet the performance criteria
after one year.

\begin{figure}[htbp]
\begin{center}
\mbox{\epsfig{figure=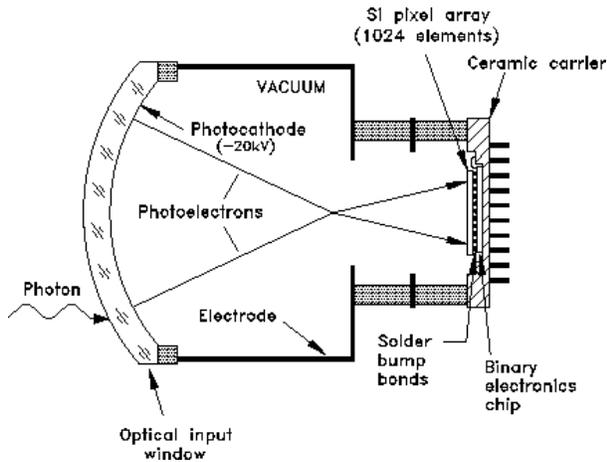,width=0.5\textwidth}}
\caption{  Schematic of the Pixel HPD, 
 illustrating photo electron trajectories.}
\label{fig:hpd}
\end{center}
\end{figure}
A schematic drawing of the HPD is shown in Figure~\ref{fig:hpd}.
A HPD has a 7\,mm thick, spherical quartz entrance window
with an S20 (multialkali) photo cathode deposited on its inner surface.
Its integrated quantum efficiency
is $\int{Q dE} \leq 0.7 \rm{eV}$.
The active element of the HPD is a silicon diode sensor 
divided into  an array of $ 32 \times 32$ pixels of size 
$0.5\,$mm$\times 0.5\,$mm.
A cross-focusing optics provides an image demagnification of 
a factor of five from the photo cathode to
the silicon diode sensor.
The operating
voltage is $-20\,$kV at the photo cathode
and the silicon sensor anode is at ground potential. 
The HPD is read out by 
a 1024 channel binary front-end chip which
is   encapsulated inside the vacuum envelope. 
The HPD has been developed together with
industry.

\begin{figure}[hbtp]
\begin{center}
\mbox{\epsfig{figure=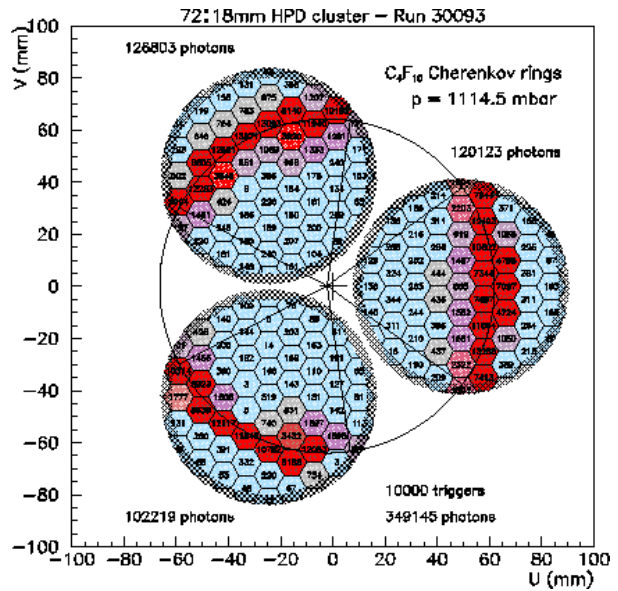,width=0.5\textwidth}}
\caption{Display of Cherenkov photons integrated over a run 
for the  three HPDs.}
\label{fig:3hpd}
\end{center}
\end{figure}
A few major results  of the test programme follow.
Three full scale prototype HPDs have been manufactured
and were fitted with a 61-pixel silicon anode. The signals
are read out externally by a Viking VA2 chip.
The three 61-pixel HPDs have been mounted onto a full-scale RICH\,1 prototype
which was operated with a $\sim 1 \rm{m}$  $\rm{C}_4\rm{F}_{10}$ radiator.
Using the CERN SPS facility this setup has been
exposed to 120~GeV/c $\pi^-$  beam. This study
tests the photon yield of the tube. Figure~\ref{fig:3hpd}
displays the integrated events of a run. The Cherenkov ring is clearly
visible and spans the three HPDs. The observed yield is in excellent agreement
with the expectations, the figure of merit is measured to be 
$202 \pm 16 \rm{cm}^{-1}$.
The 1024 channel pixel chip provides binary output signals at LHC speed.
A first design has been submitted and is currently in production. 
If the resulting chip were functional it will be encapsulated into the HPD.
An array of $3\times 3$ MaPMTs has also been evaluated in test beams
and the measurements are in excellent agreement with the expectations. 
Thus the MaPMT option meets the all the performance criteria 
if the Pixel HPD fails to meet its milestones, although with an increased cost.

Based on the results obtained for the HPD data described above
a full simulation of the RICH system has been performed. 
For the pattern reconstruction  
a global likelihood function comprising of the different
particle hypotheses for all charged tracks is maximised
by using  all the hits in RICH\,1 and RICH\,2 simultaneously.
The number of expected photo electrons for tracks with $\beta = 1$ 
is 7, 33, and 18 for the Aerogel,  $\rm{C}_4\rm{F}_{10}$, and $\rm{CF}_4$
radiator, respectively. The corresponding angular resolutions are
2.00 mrad, 1.45 mrad, and 0.58 mrad, respectively.
The performance of the system is expressed as pion-kaon separation as
a function of track momentum and 
\begin{figure}[bhtp]
\begin{center}
\mbox{\epsfig{figure=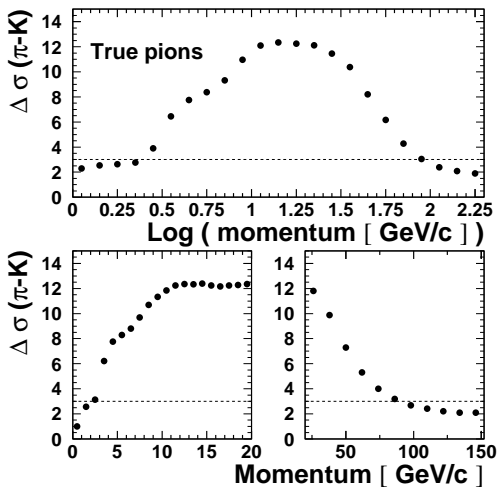,width=0.45\textwidth}}
\caption{Pion-kaon separation for logarithmic and linear momentum scales.}
\label{fig:piksep}
\end{center}
\end{figure}
is shown in Figure~\ref{fig:piksep}. This  demonstrates that
the LHCb RICH system provides $2.5 \; \sigma$ or larger pion-kaon 
separation over the momentum
range 1-150 GeV/c and    $6 \; \sigma$ or   larger separation between
3 - 60 GeV/c.

Mechanical design studies for RICH\,1 and RICH\,2 have been carried out.
The total material of the two detectors 
is 14~\% and 12~\% of a radiation length, respectively.
For RICH\,1 the beam pipe is an integral part of the 
detector vessel and a special seal has been  designed to minimize stresses
to the beam pipe and to allow for bake-outs.
The technical design report for the RICH system has been submitted. 

\section{Calorimeters}

The LHCb calorimeters provide information  for the Level-L0 trigger
and for the reconstruction 
of electrons, neutral pions and photons.
The electromagnetic calorimeter  
is of the ``shashlik''-type. Lead and scintillator tiles alternate,
the total thickness is 25 radiation lengths. 
Using wave-length shifting fibres for light collection 
the signals are read out by  photo multiplier tubes.
Keeping the number of read-out channels as a constraint
it was found that  the trigger and reconstruction efficiencies 
can be optimised by employing
three different cell sizes, 
from 40.4~mm for the inner sections to 
to 121.2~mm for the outer sections. 
\begin{figure}[htb]
\begin{center}
\mbox{\epsfig{figure=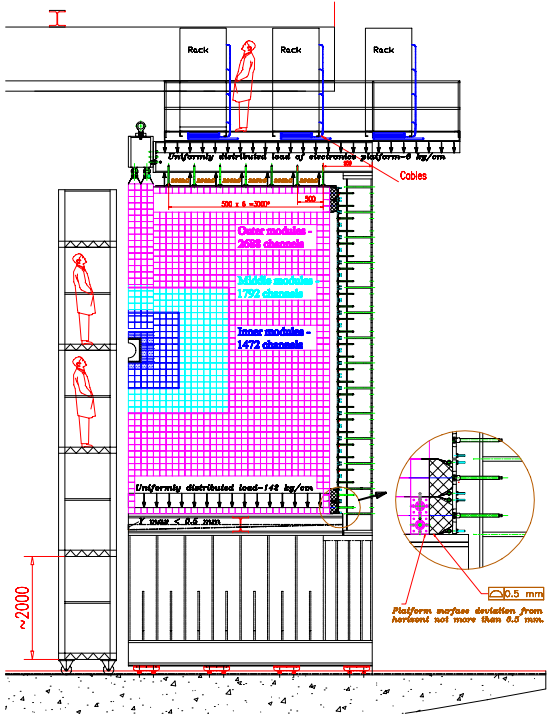,width=0.5\textwidth}}
\caption{Front view of one half of the electromagnetic calorimeter.}
\label{fig:ecal}
\end{center}
\end{figure}
In Figure~\ref{fig:ecal} we show a schematic view of a detector half.
In front of the electromagnetic calorimeter there is
a preshower detector consisting of a sandwich 
of scintillator pads - lead  - pre-shower scintillators. 
The signals are read out 
by multianode photo multiplier tubes.
This pre-shower provides electron-pion separation for the L0 trigger
and improves upon the energy resolution.
The scintillator pads allow for electron-photon separation in the 
L0 trigger.

The hadron calorimeter  comprises of alternating iron/scintillating tiles
with the tiles parallel to the beam. As in the electromagnetic
calorimeter the light is collected with wave-length shifting fibres 
and the signals are read out by photo multiplier tubes.
To facilitate the L0 trigger implementation a projective
cell geometry is used for the calorimeters.
For the hadron calorimeter the granularity is reduced
and only  two cell sizes are needed. 
The larger (smaller) one is twice (equal to) 
the projection of an outer section  electromagnetic cell.
This allows to construct the modules of the hadron calorimeter out of 
a single mechanical type which reduces complexity and cost.
The total thickness has been reduced 
from 7.3 to 5.6 nuclear interaction length.
This will introduce  a negligible loss  of resolution. 

An extensive test programme for the calorimeters has been carried out
over the last few years. 
The design value of the energy resolution for the electromagnetic calorimeter,
$ {\sigma_E}/{E} = {10 \%}/{\sqrt{E}} \oplus 1.5\%$, 
has been obtained in measurements 
using prototype detectors in beam  tests.
The inner most electromagnetic modules
have to sustain a radiation of up 0.25 Mrad/year.
We have demonstrated  by irradiating the scintillators and the wave-length
shifters up to 5 Mrad that a subsequent annealing for 175 hours
retained a total light yield such that the constant term of the resolution  
will not exceed the design value for more than 10 years of operation.
\begin{figure}[htbp]
\begin{center}
\mbox{\epsfig{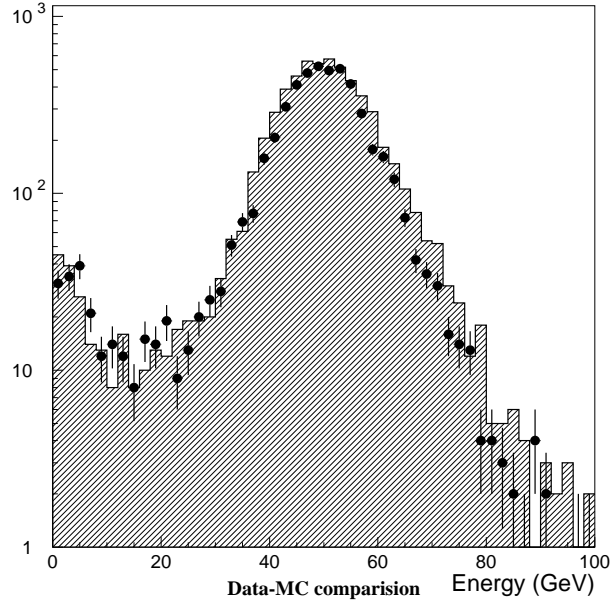}}
\caption{Energy response of a hadron calorimeter prototype 
for 50 GeV pions from test beam data (hatched) and 
and from simulation (dots).}
\label{fig:hcal}
\end{center}
\end{figure}
In Figure~\ref{fig:hcal} we plot the response of a hadron calorimeter module
to 50 GeV pions. The test beam data agree well with the simulation which
predicts an energy resolution of 
$ {\sigma_E}/{E} = {80 \%}/{\sqrt{E}} \oplus 5\%$.
The performance of the calorimeter system has also been measured 
by testing preshower, electromagnetic, and hadron calorimeter modules jointly.
The front-end electronics board has been designed and the 
Technical Design Report has been submitted. Construction of the calorimeters
will start in 2001.

\section{Trigger}

The bunch crossing frequency at the LHC is 40~MHz. The LHCb
experiment will run at a tunable\footnote{The design luminosity
of the LHC is  $10^{34} \rm cm^{-2} \rm s^{-1}$ but 
by defocusing the beam optics   it can be tuned 
to lower values at the LHCb interaction point.}
luminosity of  $2 \times 10^{32} \rm cm^{-2} \rm s^{-1}$.
This relatively low luminosity 
optimises the number of bunch crossings with one single $p p$ interaction,
thus minimising the detector occupancies and reducing the radiation levels.
It will be achieved in the first year of LHC running.

The LHCb trigger system comprises of four levels. The strategy  is to have 
a flexible multi level trigger which does not rely on a single
detector. The trigger must be stable and robust under varying
conditions. A first trigger level, L0,  
implemented in hardware with a latency of $4 \mu \rm{s}$,
reduces the event rate to 1~MHz. 
The L0 trigger accepts events containing particles
with large transverse momenta, $p_{\perp}$, which substantially
enhances the fraction of $b$-quark events.
Using the muon and the calorimeter systems
high $p_{\perp}$  muons, hadrons, and electrons, photons, and neutral pions
are selected.

The second trigger level is the L1 vertex trigger which must reduce 
the rate to 40~kHz.
Taking advantage of the vertex geometry design
track finding is firstly done using $r$-strips only.
A primary vertex is reconstructed and secondary tracks are selected.
Secondly the information from the $\varphi$-strips 
is added to select events with detached secondary vertices.

Trigger levels L2 and L3 are software based processor farms. In L2
the information from all the tracking systems is accessible whereas 
for L3 the full event information will be used. 
The data acquisition system stores events at a rate of 200~Hz.
Typical trigger efficiencies are around 30~\%.
Simulations show that the hadron trigger is important.
The tagging of the decay flavour of the $\mathrm{B}$ meson is based on
decay products of the other $\mathrm{B}$ hadron:
either the charge of the muons and electrons 
from semileptonic decays of the other $\mathrm{B}$ hadron or
the charge of the kaons stemming from the decay chain 
$b \, \rightarrow \, c \, \rightarrow \, s$.
The total tagging  efficiency  is $\varepsilon = 40 \%$,
the mistag rate is $w = 30 \%$, leading to a effective
dilution factor of $D = \varepsilon (1 -2w)^2 = 6.4\%$.

\section{Physics Performance}

\begin{figure}[!t]
\begin{center}
\mbox{\epsfig{figure=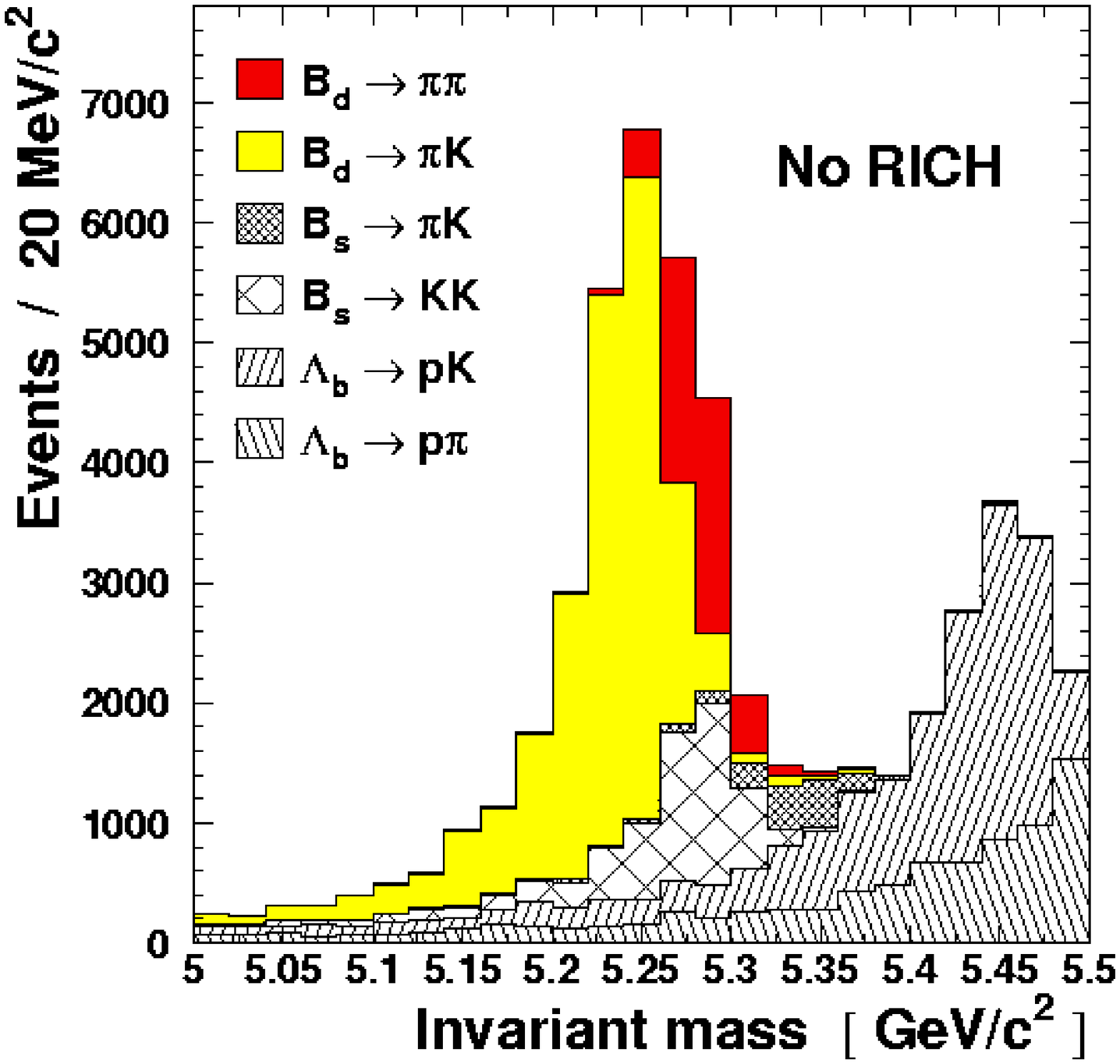,width=0.25\textwidth}
        \epsfig{figure=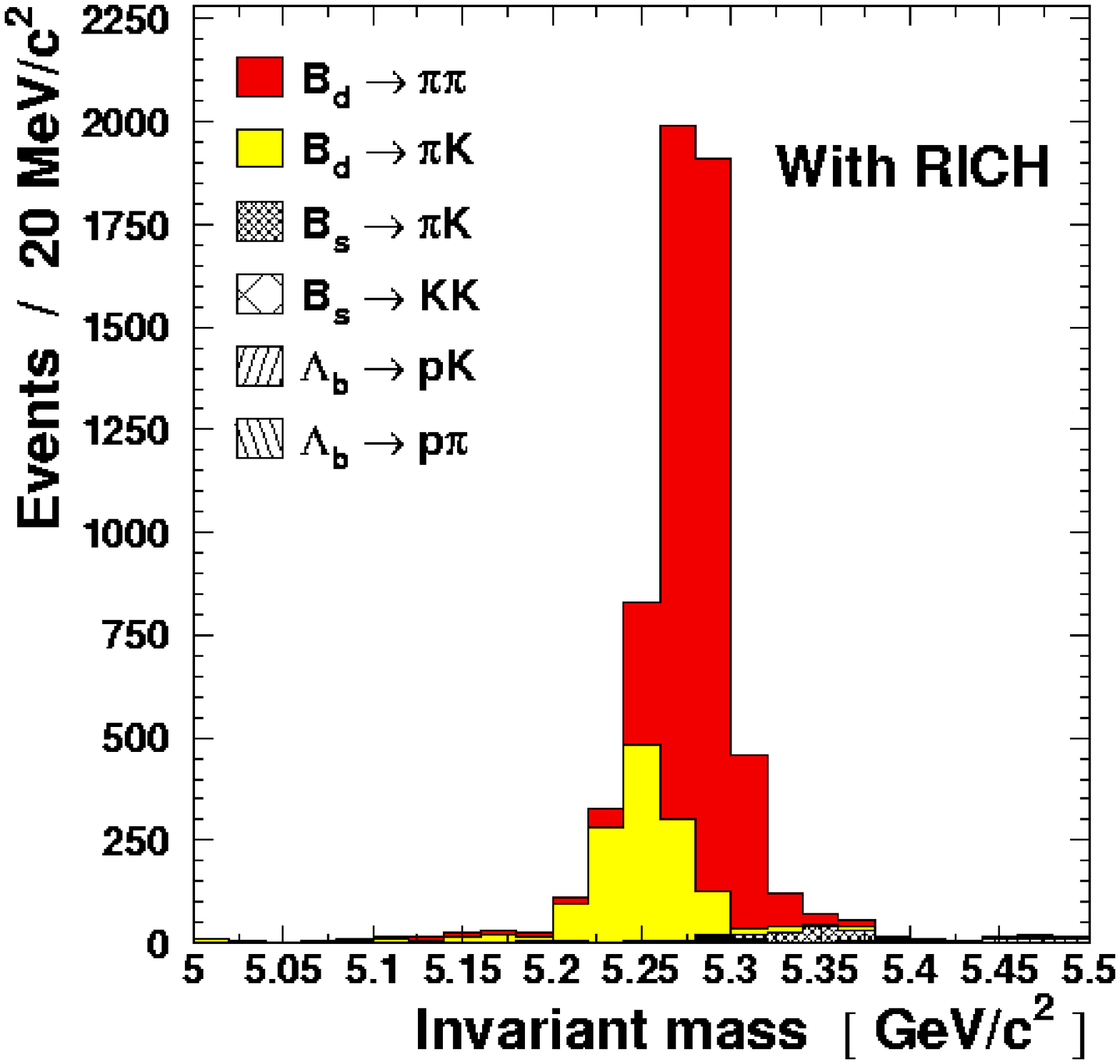,width=0.25\textwidth}}
\caption{Mass spectrum of  $\mathrm{B^0_d  \, \rightarrow \, \pi^- \pi^+} $ candidates
before  and after particle identification.}
\label{fig:bpipi}
\end{center}
\end{figure}

Here we describe the expected LHCb performance for a few selected channels. 
The  trigger efficiencies, tagged event yields and  
LHCb sensitivities are  compiled in Table~\ref{tab:performance}. 
The excellent performance of the RICH system is demonstrated
on simulated $\mathrm{B}$ hadrons decaying in two tracks.
On the left-hand side of Figure~\ref{fig:bpipi}
we plot  
the invariant mass of the two reconstructed  charged tracks without particle
identification.  The signal for the decay mode 
 $\mathrm{B^0_d  \, \rightarrow \, \pi^- \pi^+} $ which 
is sensitive to the CP violating
unitarity angle $\alpha$ is sitting on the shoulder 
of the larger background mode
 $\mathrm{B^0_d  \, \rightarrow K^- \pi^+} $. 
Using the RICH system retains the signal with a high efficiency, 
whereas the  background modes are eliminated or  greatly reduced.
%
%
%
This clear separation is even more important since some of the background 
modes can have CP violation as well. LHCb expects to reconstruct 4900 tagged
signal events in one year which corresponds to a sensitivity of 
$\sigma_{\alpha} = 2 - 5^\circ$.~\footnote{This results assumes that the 
contribution from the penguin diagram to 
 $\mathrm{B^0_d  \, \rightarrow \, \pi^- \pi^+} $ can be extracted
separately.}

\begin{figure}[bhtp]
\begin{center}
\mbox{\epsfig{figure=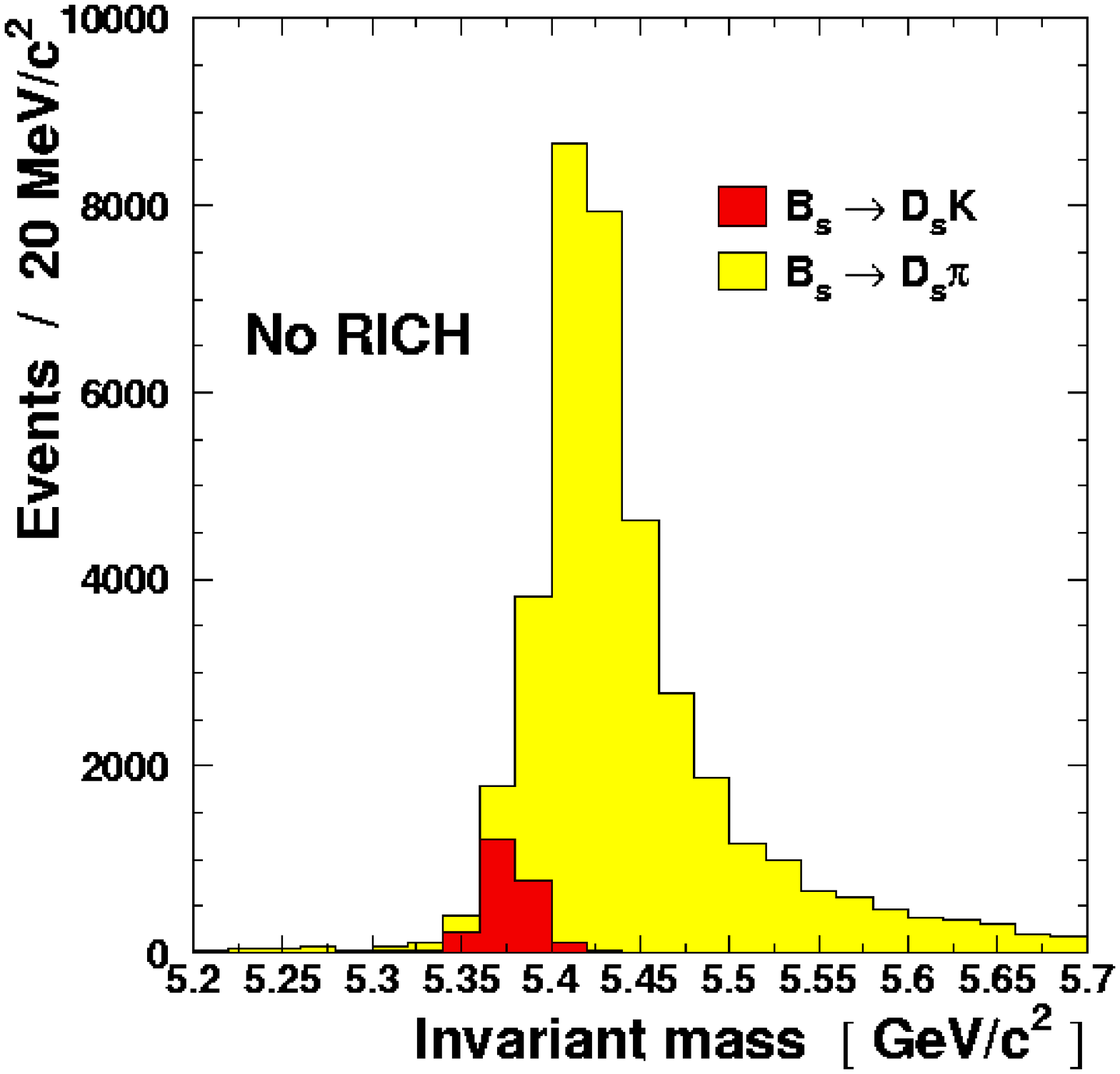,width=0.25\textwidth}
        \epsfig{figure=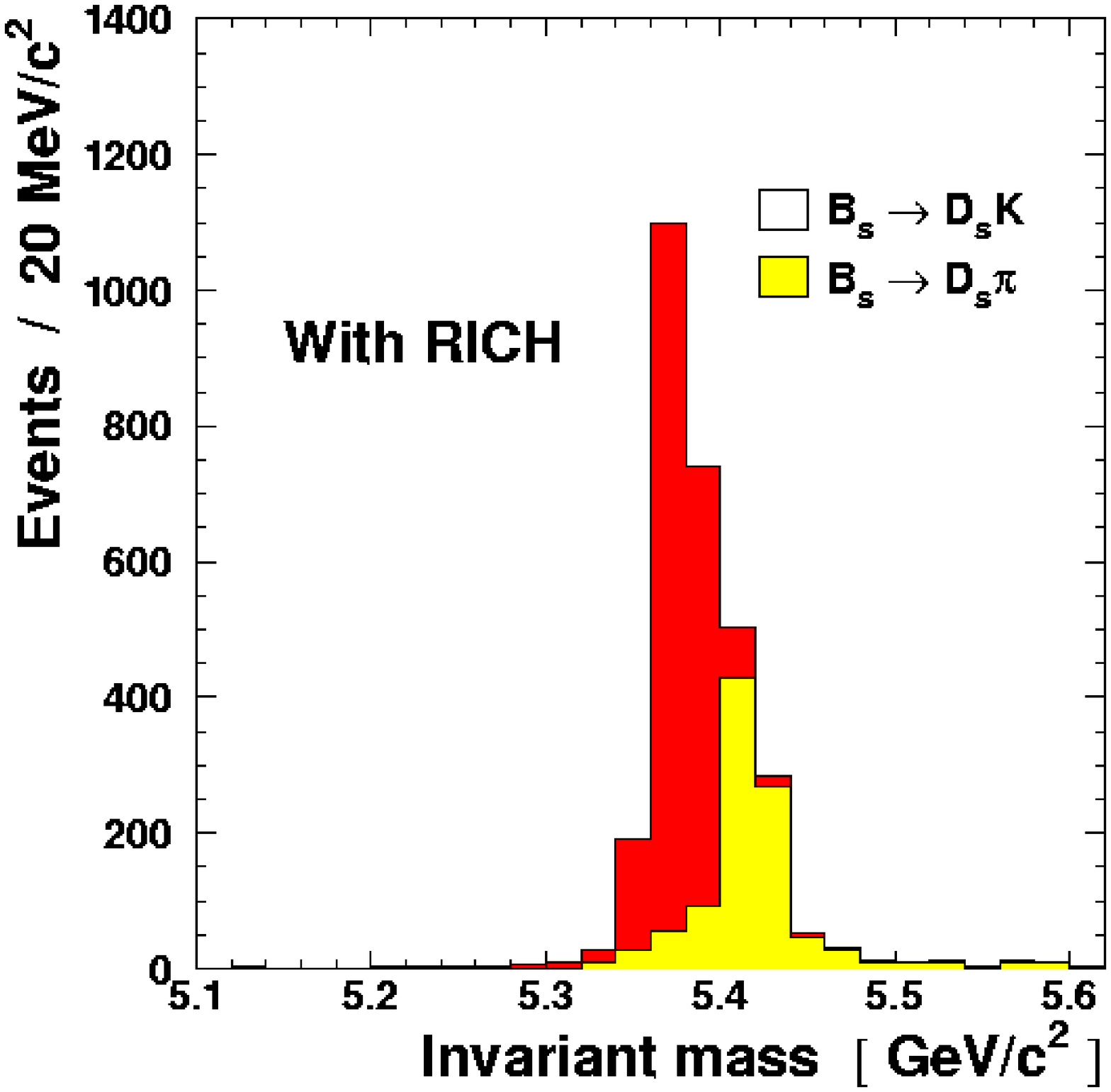,width=0.25\textwidth}}
\caption{Mass spectrum of $\mathrm {B^0_s \, \rightarrow \, D_s^{\mp} K^{\pm}}$
candidates before  and after particle identification.}
\label{fig:bsdsk}
\end{center}
\end{figure}
In Figure~\ref{fig:bsdsk} we plot
the invariant mass distribution for the channel
$\mathrm {B^0_s \, \rightarrow \, D_s^{\mp} K^{\pm}}$. 
The two charge conjugate decays
are caused by two tree level diagrams which are of the same order of magnitude.
Large CP violating effects are possible in the interplay
of $\mathrm{B_s \bar B_s}$ mixing and  decay.
Without particle identification the signal is dwarfed by the Cabibbo
favoured mode 
$\mathrm {B^0_s \, \rightarrow \, D_s^{\mp} \pi^{\pm}}$.
After applying the RICH selection 
the signal dominates. This  is crucial
for measuring the four time dependent rate asymmetries which  allow a determination 
of the unitarity angle $\gamma - 2 \delta \gamma $.
About 2400 tagged events are collected in one year of data taking 
and a measurement precision of $\sigma_{\gamma - 2 \delta \gamma} = 
6 - 14 ^\circ$ is expected.

\begin{figure}[htbp]
\begin{center}
\mbox{\epsfig{figure=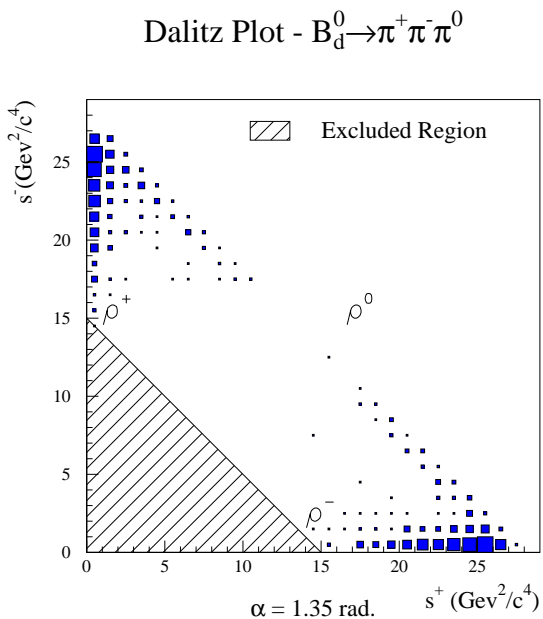,width=0.50\textwidth}}
\caption{The Dalitz plot for $\mathrm{B^0_d  \, \rightarrow \, \rho \pi} $.}
\label{fig:brhopi}
\end{center}
\end{figure}

The decay channel
 $\mathrm{B^0_d  \, \rightarrow \, \rho \pi} $ 
is sensitive to the unitarity angle $\alpha$. 
In Figure~\ref{fig:brhopi} we show a Dalitz plot for reconstructed
signal events. A time dependent analysis allows to
extract not only $\alpha$ but also the strength of both the contributing
tree  and penguin amplitudes. About 1300 tagged events are reconstructed
in one year LHCb running yielding a sensitivity of
$\sigma_{\alpha} =  3 - 6^\circ$.

\begin{table*}
\begin{center}
\begin{tabular}{|c|c||ccc|} \hline
Parameter & Channel & Trigger efficiency & Event yield & Sensitivity \\ \hline
$\alpha$  &  $\mathrm{B^0_d \,\rightarrow \, \pi^+ \pi^-} ^a$  
& $30\%$  & $4.9$k & $2-5^\circ$ \\
          &  $\mathrm {B^0_d \, \rightarrow \, \rho \pi}$ 
& $20\%$  & $1.3$k & $3-6^\circ$ \\ 
$ 2\beta + \gamma$  & $\mathrm {B^0_d \, \rightarrow \, D^\star \pi}$ 
& $33\%$  & $340$k & $>11^\circ$ \\ \hline
$\beta$   & $\mathrm{B^0_d \,\rightarrow \, J/\psi K^0_s}$ 
& $36\%$ & $37$k & $0.6^\circ$ \\ \hline
$\gamma - 2 \delta \gamma$ & $\mathrm{B^0_s \, \rightarrow \, D_s K}$ 
& $28\%$ &  $2.4$k & $6-14^\circ$ \\
$\gamma$  & $\mathrm{B^0_d \,\rightarrow \, D^0 K^{\star 0}}$ 
& $21\%$ &  $0.4$k & $10^\circ$ \\ \hline
$\delta \gamma$ & $\mathrm{B^0_s \, \rightarrow \, J/\psi \phi}$  
& $38\%$  & $44$k & $0.6^\circ$ \\  \hline
$x_s$     & $\mathrm{B^0_s \, \rightarrow \, D_s \pi}$ 
& $28\%$ & $120$k & up to $x_s \, \sim \, 75$ \\ \hline
Branching ratio & $\mathrm{B^0_s\,\rightarrow \mu^+ \mu^-}$
& $95\%$  & $11$    &      $< 2 \times 10^{-9}$       \\ 
          & $\mathrm{B^0_d\,\rightarrow K^{\star 0} \gamma}$
& $8\%$   & $26$k   &      --      \\ 
          & $\mathrm{B^0_d\,\rightarrow K^{\star 0} \mu^+ \mu^-}$
& --      & $4.5$k   &      --      \\ \hline
\end{tabular}
\end{center}
\caption[]{LHCb performance for selected benchmark channels 
for one year of LHC operation.}
\label{tab:performance}
\end{table*}

New strategies for measuring CKM angles in direct CP violation
using the modes 
$\mathrm{B^0_d  \, \rightarrow \, K \pi     } $~\cite{ref:buchalla},
$\mathrm{B^0_{d,s}  \, \rightarrow \, \pi \pi,  K  K} $~\cite{ref:gronau}, and
$\mathrm{B^0_{d,s}  \, \rightarrow \, D^+_{d,s}D^{-}_{d,s}} $
are also being developed.
Complementary searches for new physics can be performed outside the unitarity
triangle by searching for CP violation in channels which are expected
to have very low Standard Model asymmetries. Examples 
are decays generated by Penguin diagrams only. LHCb has studied
the sensitivity to $\mathrm{B^0_d  \, \rightarrow \, K^{*0} \gamma} $,
and to $\mathrm{B^0_d  \, \rightarrow \, K^{*0} \mu^+ \mu^-} $
where the forward-backward asymmetry is sensitive to 
contributions  from MSSM models as discussed in another 
presentation~\cite{ref:ali}. 
Alternatively channels may be searched for which have extremely 
low, but cleanly predicted Standard Model branching ratios. For the decay
$\mathrm{B^0_s  \, \rightarrow \, \mu^+ \mu^-} $ the SM predicts 
a  branching ratio of $3.7 \times 10^{-9}$  and yet LHCb expects
to see about 11 events of this  decay  within one year of data taking.

\section{Conclusions}

The LHCb experiment is progressing rapidly since the Technical Proposal.
Major technology choices have been made.
A normal conductive coil has been chosen for the magnet.
The pixel HPD has been selected as the baseline photo detector for the RICH system.
The Technical Design Reports for the magnet has been approved, and the TDRs
have just been submitted for the RICH and calorimeters. Other
subsystems such as vertex detector are well on track.
Some subsystems, e.g. the calorimeters, will enter into the construction
phase in early 2001.
The trigger system has demonstrated its robustness and the experiment 
will be able to collect data from the start of the LHC.
The physics performance studies have been extended and 
show that there will be a long and healthy physics programme for the 
LHCb experiment.

\section*{Acknowledgments}

With great pleasure I thank my LHCb colleagues for their help in
the preparation of this contribution.

\end{document}